# PHASECOHERENT TRANSPORT IN PbTe WIDE PARABOLIC QUANTUM WELLS

J. OSWALD, G. SPAN, A. HOMER, G. HEIGL, P. GANITZER,
Institute of Physics, University of Leoben, A-8700 Leoben, Austria
E-mail: oswald@unileoben.ac.at

D.K. MAUDE, J.C. PORTAL
High Magnetic Field Laboratory, CNRS, BP 166 Grenoble, France

Conductance fluctuations have been observed in macroscopic, quasi 3D PbTe wide quantum wells. A significant increase of the correlation field occurs in a temperature range from 40 mK to 1.2K. At the same time the fluctuation amplitude stays near $e^2/h$ although the lateral sample size is two orders of magnitude larger than any typical length scale of diffusive electron transport. We interpret this behavior in terms of phasecoherent electrontransport, which takes advantage of a dramatic enhancement of the phasecoherence length of electrons in edgechannels and edge channel loops in the bulk region.

## 1 Introduction

A study of the intermediate regime between 2-dimensional (2D) and 3-dimensional (3D) electronic systems has attracted a lot of interest in the past [1]. However, the realization of such so called „wide parabolic quantum wells" (WPQW) with a flat potential in the electron channel requires a parabolic bare potential, which is difficult to obtain on the basis of GaAs/Al$_x$Ga$_{1-x}$As heterostructures by MBE growth. WPQWs can be made much more easily using the n-i-p-i concept with Lead Telluride (PbTe) and have been recently successfully realized and investigated [2, 3]. The main facts are as follows: The quantum Hall effect is suppressed and replaced by conductance fluctuations (CF) of universal amplitude $e^2/h$ [4,5,6]. The magnetotransport is dominated by edge channel conduction with permanent back scattering, which gives rise to a linear increase of $R_{xx}$ with magnetic field [7]. The CF are explained to result from fluctuations of this back scattering process [5,6].

## 2 Experimental results

The fluctuations appear basically superimposed on a linearly increasing magneto resistance $R_{xx}$. After obtaining the fluctuations in the resistance $\Delta R$ by subtracting the linear background resistance, the conductance fluctuations have been calculated from $\Delta R$ by first order expansion according to $\Delta G = (\Delta R/R^2)$. Further experimental details can be obtained from [2,4,5,6]. The left part of Fig.1 shows rms. value and correlation field of the CF for 3 different samples versus temperature. There can be found a significant increase of $B_c$ with temperature without indicating any saturation with decreasing temperature.

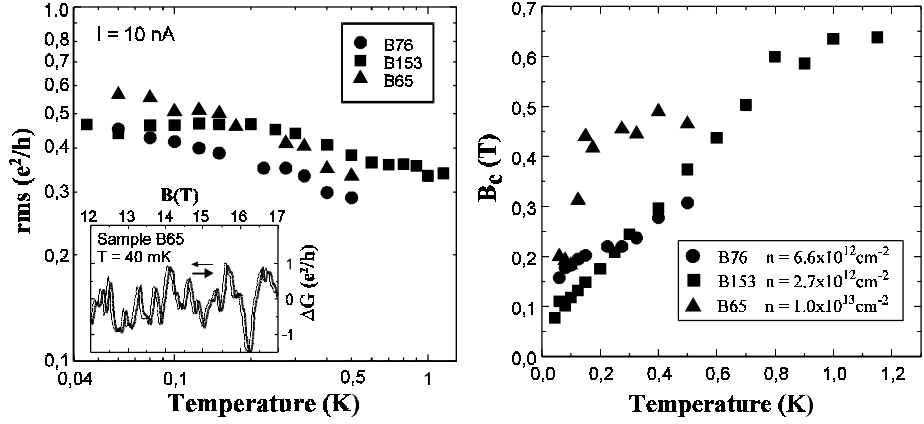

**Figure 1 Left:** rms. value of the conductance fluctuations versus temperature for 3 different samples. The insert shows typical CF data extracted from $R_{xx}$. **Right:** Correlation field $B_C$ versus temperature.

The electron mobility is well above $10^5 cm^2 V^{-1} s^{-1}$, the distance between the voltage probes is 200μm and the width of the electron channel is 330nm. The samples differ in the electron density of the channel, as indicated in Fig.1.

## 3 Discussion

The large fluctuation amplitude suggests phasecoherent transport on the entire macroscopic sample while the consideration of the correlation field $B_C$ results in a typical area between phasecoherent paths of about 100 mm diameter. The main idea of our interpretation is that the absence of back scattering of edge electrons directly in the same edge channel leads also to a drastic enhancement of the phasecoherence length as compared to free electrons. The same argument applies to electrons in closed EC-loops of the bulk region, which represent basically equipotential lines of the native lateral potential fluctuations. Since pure resonant tunneling between different loops will not be phase destroying, we expect that it is possible to have electrons traveling phase coherently over macroscopic distances even in the bulk region. The basic idea of this mechanism is schematically shown in Fig.2.

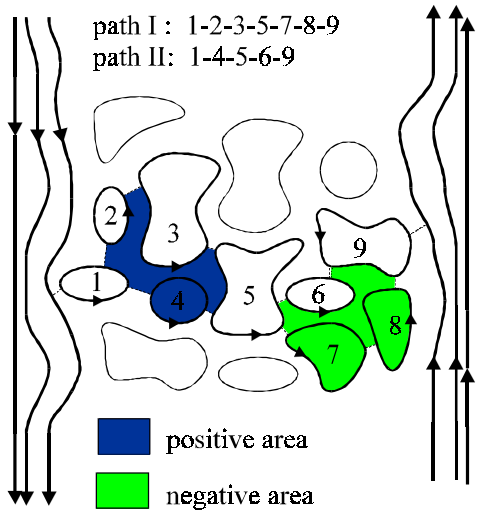

**Figure 2:** Schematic representation of phase-coherent electron transport across the bulk region by resonant tunneling between magnetic bound states. A typical path follows a loop in the proper direction until it leaves somewhere by tunneling and continues in the next loop. Every loop potentially allows a crossing of different paths analogous to a "circular traffic crossing". In this example the two arbitrarily selected paths cross each other at loop 5, resulting in a partial cancellation of the total enclosed area.

As explained by Fig.2, a complex network of phasecoherent paths for the transmission of electrons across the bulk region can develop. The dominating network character of the paths must in general lead to a significant cancellation of the enclosed flux between any two paths, which reduces the effective area for the associated Aharonov Bohm (AB) oscillations. In this way we expect a contribution of large and small AB loops, which are composed by the statistically arranged paths of the network.

The next important point is the temperature induced smearing-out of the quantum interference. The general effect behind is that edge electrons with slightly different energy acquire also a slightly different phase if traveling around the same loop. Consequently a thermaly induced uncertainty in energy will cause an uncertainty in phase. This smearing-out was already analyzed in the past for the case of the transmission across a single magnetic bound state by considering the Bohr-Sommerfeld spectrum of an idealized magnetic bound state [8].

$$\Delta \varepsilon = \left(\frac{h}{e}\right) \frac{e \cdot E_r}{2\pi \cdot r \cdot B} \qquad (1)$$

Eqn.1 describes a circular magnetic bound state where $\Delta\varepsilon$ is the separation of the energy levels, $r$ is the radius of the loop, $E_r$ is the electric field which binds the electrons to the potential hill and $B$ is the magnetic field. The basic criterion, which leads to the Bohr-Sommerfeld condition, is constructive self-interference of the bound "edge electrons". Therefore the spectrum of Eqn.1 can be understood in the following way: If the energy of a bound electron is increased, also the acquired phase changes and the electron will fit only into the next energy level if the additional phase is exactly $2\pi$ in order to get again constructive self-interference. Considering thermal smearing-out, $\Delta\varepsilon$ will be of the order of the thermal energy $k_B T$. Since $\Delta\varepsilon$ depends on the reciprocal radius $r^{-1}$ of the loop, an increase of temperature will lead first to a disappearance of the fast AB-oscillations from large loops while leaving the slowly oscillating contributions of the smaller loops. This must result in an increase of the correlation field $B_c$ with temperature.

Looking at Fig.1 right, $B_C$ increases indeed significantly with temperature. We use Eqn.1 and estimate the temperature which corresponds to a typical correlation field of $B_C = 0.15$ T. The corresponding mean radius of the loops is 90 nm. According to estimates based on [9] we get a value for the mean electric field of the potential fluctuations of about $E_r = 1.6x10^4$ V/m. Since the fluctuations can be analyzed in the magnetic field range above 10 Tesla only, a typical magnetic field of $B = 10\ T$ is used in the calculation. Now we can calculate $\Delta\varepsilon$ from Eqn.1 and get $\Delta\varepsilon = 12$ µeV. This value corresponds to a temperature of T = 135 mK. This temperature agrees perfectly with the experimental data of Fig.1 (right) where the corresponding temperature is found to be about 100 mK.

**Acknowledgments**

Support by FWF Austria under project P10510-NAW and through the TMR Programme of the European Community under contract ERBFMGECT950077.